\documentclass[prl,showpacs,amsmath,amssymb,superscriptaddress,twocolumn]{revtex4-1}
\usepackage{graphicx}
\usepackage{dcolumn}
\usepackage{color}
\usepackage{hyperref}

\newcommand*{\figurewidth}{0.9\columnwidth}

\begin{document}

\title{Prominent role of multi-electron processes in K-shell double and triple photodetachment of oxygen anions}

 \author{S.~Schippers}
 \affiliation{I.~Physikalisches Institut, Justus-Liebig-Universit\"{a}t Gie{\ss}en, Heinrich-Buff-Ring 16, 35392 Gie{\ss}en, Germany}
 \email[]{stefan.schippers@physik.uni-giessen.de}

 \author{R.~Beerwerth}
 \affiliation{Helmholtz-Institut Jena,  Fr{\"o}belstieg 3, 07743 Jena, Germany}
 \affiliation{Theoretisch-Physikalisches Institut, Friedrich-Schiller-Universit\"{a}t Jena, 07743 Jena, Germany}

 \author{L.~Abrok}
 \affiliation{Institute of Nuclear Research of the Hungarian Academy of Sciences, Debrecen, P.O. Box 51, H-4001, Hungary}

 \author{S.~Bari}
 \affiliation{European XFEL GmbH, Holzkoppel 4, 22869 Schenefeld, Germany}
 \affiliation{FS-SCS, DESY,  Notkestra{\ss}e 85, 22607 Hamburg, Germany}

  \author{T.~Buhr}
 \affiliation{Institut f\"{u}r Atom- und Molek\"{u}lphysik, Justus-Liebig-Universit\"{a}t Gie{\ss}en, Leihgesterner Weg 217, 35392 Gie{\ss}en, Germany}

  \author{M.~Martins}
 \affiliation{Institut f\"{u}r Experimentalphysik, Universit\"{a}t Hamburg, Luruper Chaussee 149, 22761 Hamburg, Germany}

 \author{S.~Ricz}
 \affiliation{Institute of Nuclear Research of the Hungarian Academy of Sciences, Debrecen, P.O. Box 51, H-4001, Hungary}

 \author{J.~Viefhaus}
 \affiliation{FS-PE, DESY,  Notkestra{\ss}e 85, 22607 Hamburg, Germany}

 \author{S.~Fritzsche}
 \affiliation{Helmholtz-Institut Jena, Fr{\"o}belstieg 3, 07743 Jena, Germany}
 \affiliation{Theoretisch-Physikalisches Institut, Friedrich-Schiller-Universit\"{a}t Jena, 07743 Jena, Germany}

 \author{A.~M\"uller}
 \affiliation{Institut f\"{u}r Atom- und Molek\"{u}lphysik, Justus-Liebig-Universit\"{a}t Gie{\ss}en, Leihgesterner Weg 217, 35392 Gie{\ss}en, Germany}

\date{\today}

\begin{abstract}
The photon-ion merged-beams technique was used at a synchrotron light source for measuring absolute cross sections of double and triple photodetachment of O$^{-}$ ions. The experimental photon energy range of 524--543 eV comprised the threshold for K-shell ionization. Using resolving powers of up to 13000, the position, strength and width of the below-threshold $1s\,2s^2\,2p^6\;^2S$ resonance as well as the positions of the  $1s\,2s^2\,2p^5\;^3P$ and $1s\,2s^2\,2p^5\;^1P$ thresholds for K-shell ionization were determined with high-precision. In addition, systematically enlarged multi-configuration Dirac-Fock calculations have been performed for the resonant detachment cross sections. Results from these ab-initio computations agree very well with the measurements for the widths and branching fractions for double and triple detachment, if \emph{double} shake-up (and -down) of the valence electrons and the rearrangement of the electron density is taken into account. For the absolute cross sections, however, a previously found discrepancy between measurements and theory is confirmed.
\end{abstract}


\maketitle

Negative atomic ions play an important role in low-temperature plasmas such as the upper atmosphere or the interstellar medium \cite{Andersen2004b} and also in technical applications. For example, in the context of antihydrogen production, it has been proposed to use an ensemble of laser-cooled anions as a coolant for antiprotons \cite{Kellerbauer2006}. Negative ions are fundamentally different from neutral atoms or positive ions since the extra electron in a negative ion is not only bound by the long-range Coulomb interaction with the atomic nucleus but, more importantly, also by a short-range attractive force due to the polarization of the atomic core. The accurate theoretical description of these ions still challenges the state-of-the-art quantum computations although the numbers of their bound states are generally finite. The low-excitation levels of negative ions are readily accessible by laser spectroscopy (see, e.g., \cite{Lindahl2012,Walter2014,Baeckstroem2015a,Jordan2015}). Therefore, this technique has been a prime source of experimental information about the mutual interactions among the valence electrons.

A sensitive tool for studying the interactions between the valence and the core electrons is inner-shell ionization of negative ions \cite{Kjeldsen2001a,Berrah2001}. Here, we apply the photon-ion merged-beams technique (see \cite{Schippers2016} for a recent overview) to determine the absolute cross sections for double and triple ionization (detachment) of oxygen anions in the photon energy range 524--543~eV. In this energy range, a $K$-shell vacancy may be produced either by direct ionization of an initial $1s$ electron or via the formation of a resonance state by exciting one $1s$ electron to a higher shell such as $2p$. In either case, the $K$-vacancy decays subsequently by a cascade of radiative and nonradiative processes leading to a distribution of final charge states with O$^{+}$ and O$^{2+}$ as the most prominent charged reaction products. To test and better understand the theoretical prediction of such cascades, we performed extremely comprehensive quantum calculations for the formation of the intermediate resonances and for the complex deexcitation pathways at a level of detail never attempted before. In particular, all resonance parameters (resonance energy, natural line width, and  strength) of the $1s\,2s^2\,2p^6$ photoionization resonance as well as the (absolute) cross sections and ion yields are determined independently by experiment and multi-configuration Dirac-Fock (MCDF) calculations.

Previous experimental studies of double detachment via inner K-shell excitation/ionization were performed for several anions lighter than O$^-$, i.e.,
for Li$^-$ \cite{Kjeldsen2001a,Berrah2001},  B$^-$ \cite{Berrah2007a}, and C$^-$ \cite{Gibson2003a,Walter2006a}. A recent review of related computations is provided in Ref.~\cite{Gorczyca2004}. Measured O$^+$ ion yields from double detachment of O$^-$ were presented at a conference \cite{Gibson2012}. Compared to this preliminary investigation, the present cross-section data span a much wider range of photon energies, are on an absolute scale, and were obtained with a better energy resolution. In addition, triple detachment was measured. We also note that single and double detachment of O$^{-}$ was investigated recently with a free electron laser at a fixed photon energy \cite{Harbo2012}. So far, no experimental results on $K$-shell detachment were published for anions heavier than O$^-$. In addition, the present study comprises the first experiments where, apart from the double detachment, also  the triple detachment via $K$-shell excitation and ionization of an anion is explored and analyzed in detail.

\begin{figure}
\centering{\includegraphics[width=\figurewidth]{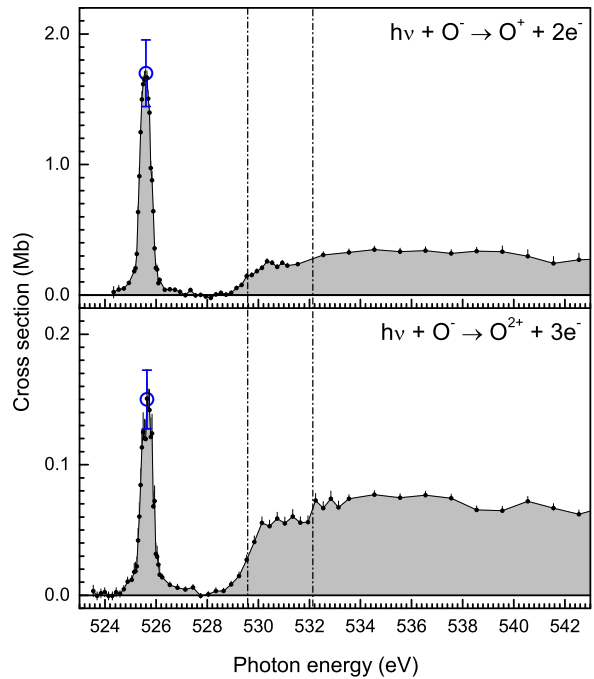}}
\caption{\label{fig:scan}Experimental cross sections (shaded curves and small symbols with statistical error bars) for double (upper panel) and triple (lower panel) photodetachment of O$^-$. The large symbols are the absolutely measured cross sections. The corresponding error bars include the $\pm15\%$ systematic uncertainty of the experimental cross section scale. The vertical dash-dotted lines mark the $^{3\!}P$ and $^{1\!}P$ K-shell ionization thresholds, respectively.}
\end{figure}

For the present experiment the photon-ion merged-beams technique was employed using the permanently installed end station PIPE \cite{Schippers2014} at the "Variable Polarization XUV Beamline" (P04) \cite{Viefhaus2013} of the  synchrotron light source PETRA\,III at DESY in Hamburg, Germany.  Negatively charged oxygen ions were produced from a gas mixture of O$_2$ and He in an electron-cyclotron resonance (ECR) ion source. After acceleration to an ion energy of 6 keV, mass/charge selection was applied by passing the ion beam through a double focussing dipole magnet. Subsequently, the ions were electrostatically guided onto the photon beam axis. The primary ion current in the merged-beams interaction region ($\sim 1.7$~m length) ranged from  50~pA to 2~nA depending on ion-beam collimation. The photon flux was measured with a calibrated photo diode. It reached $3\times10^{13}$~s$^{-1}$ at a bandwidth of 360 meV in the photon energy range of the present experiment. A second magnet, behind the interaction region, separated the O$^+$ and O$^{2+}$ product ions from the primary beam and from neutral reaction products. This demerging magnet directed the primary ions  into a Faraday cup, and the product ions into a single particle detector with a detection efficiency of practically 100\% \cite{Rinn1982}. This value was verified for 6-keV O$^+$ ions in a separate setup with an identical detector. Relative cross sections for photodetachment were obtained by normalizing the measured product-ion count rates on photon flux and ion current. The relative cross sections were put on an absolute scale by additionally measuring the spatial overlap of photon and ion beams \cite{Schippers2014}. Because this procedure is rather time consuming it has been carried out only for one photon energy, and the relative cross sections were scaled to the absolute data points as shown in Fig.~\ref{fig:scan}. This procedure is valid since the position of the photon beam did not change significantly with photon energy over the rather narrow energy range of the present experiment. The systematic uncertainty of the cross section scale is $\pm15\%$. The photon energy scale was calibrated with an uncertainty of  $\pm 0.1$~eV by remeasuring the $1s\to \pi^*$ resonance in O$_2$ at $531.06\pm0.09$~eV \cite{Prince2003b}. After this calibration of the nominal photon energy scale a Doppler correction was applied for taking the velocity of the O$^-$ ions into account \cite{Schippers2014}.

Figure~\ref{fig:scan} displays the measured cross sections for double (upper panel) and triple (lower panel) photodetachment of O$^-$. Both cross sections exhibit a similar dependence on the photon energy. A prominent resonance occurs at an energy of 525.6~eV, below the lowest threshold for direct detachment of a K-shell electron at approximately 529.6~eV. At about 535 eV, the cross sections rise to maximum values of 0.35~Mb and 0.077~Mb for double and triple detachment, respectively. At higher energies the cross sections exhibit a slow decrease as is expected for a direct ionization process of a $K$-shell electron (in this case) into the continuum. For triple detachment, the relative contribution of the direct process, as compared to the strength of the resonant process, is larger than for double detachment.

The cross section for the direct removal of a $K$-shell electron can be expected to be largely insensitive to the number of $L$-shell electrons. Indeed, the sum of our cross sections for double and triple detachment of O$^-$ at 535~eV amounts to $0.43 \pm 0.07$~Mb and agrees well with the cross section for nonresonant K-shell absorption in neutral oxygen of 0.5 Mb \cite{McLaughlin2013}. Similar values were measured for photoabsorption of O$^+$ and O$^{2+}$ \cite{Bizau2015}. In this comparison it is assumed that direct $K$-shell ionization is followed by one or more Auger decays rather than radiative transitions and that, therefore, the absorption cross section can be well approximated  by the sum of the cross sections for double and triple detachment.

\begin{figure}
\centering{\includegraphics[width=\figurewidth]{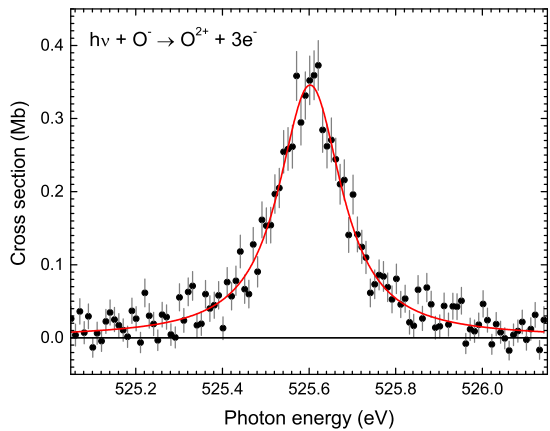}}
\caption{\label{fig:resonance} High-resolution scan (symbols) of the $1s\,2s^2\,2p^6\;^2S$  resonance in the triple detachment channel. The  full line is the result of a fit of a Voigt line-profile to the experimental data points.}
\end{figure}

The lowest threshold for K-shell ionization is associated with the $1s\,2s^2\,2p^5\;^{3\!}P_2$ $K$-vacancy level. The next higher thresholds are the $^{3\!}P_1$ and $^{3\!}P_0$ thresholds at calculated energies of 0.035 and 0.064 eV  above the $^{3\!}P_2$ levels. This small splitting is not resolved in the data of Fig.~\ref{fig:scan} due to the much larger energy spread $\Delta E =  0.36$~eV of the incident photons. However, the next higher threshold, i.\,e.,  the $1s\,2s^2\,2p^5\;^{1\!}P_1$  threshold (2.87 eV above the $^{3\!}P_0$ threshold) is clearly observed in the triple-detachment channel. No hint of this threshold was seen by Gibson et al.~\cite{Gibson2012} probably because an even larger photon energy spread was used in their experiment. For a more accurate determination of the threshold energies we have fitted step functions that were convoluted with a gaussian to the experimental data,  and this gives rise to the threshold energies $529.56\pm 0.07$~eV and $532.13\pm 0.40$~eV for the $1s\,2s^2\,2p^5\;^{3\!}P$ and $1s\,2s^2\,2p^5\;^{1\!}P$ thresholds, respectively. The given uncertainties result from the fits and do not include the 0.1~eV systematic uncertainty of our photon energy scale.

The prominent resonance at 525.6~eV is associated with the $1s \to 2p$ photoexcitation of the O$^-$($1s^2\,2s^2\,2p^5\;^2P$) ground term.
A Voigt line-profile has been fitted to the high-resolution measurement ($\Delta E\approx 40$~meV) as displayed in Fig.~\ref{fig:resonance}. This allows to experimentally extract the natural line width $\Gamma = 164 \pm 14$~meV that corresponds to a lifetime $\tau=\hbar/\Gamma = 4.0\pm0.3$~fs. The same fit yields a resonance energy of $525.603\pm0.003$~eV ($\pm$0.1~eV systematic uncertainty). Similarly, Voigt line-profile fits to the experimental cross sections from Fig.~\ref{fig:scan} yield the experimental resonance strengths in the double and triple photodetachment channels of $0.95\pm0.05$ and $0.092\pm0.005$~Mb~eV ($\pm15\%$ systematic uncertainty), respectively. The systematic uncertainty is associated with the factor that converts the relative to an absolute cross-section scale. This factor is the same for double and triple detachment, and, thus, its $\pm15$\% systematic uncertainty cancels for the ratio $10.3\pm0.8$ of the two resonance strengths.

Understanding the results of our experiment with its high-resolution observation of individual multi-electron detachment channels up to three-electron ejection presents a very formidable challenge for ab-initio theory. The present treatment involves a level of complexity that has never been envoked before. Previous computations focussed on one or very few specific decay paths that contribute to the release of a given number of electrons (e.\,g., \cite{Shimizu2000}). In our treatment we account for (almost) all of these paths, including all the low-lying shake-up (-down) transitions.  Additional challenges are posed by starting from a negative ion. We found that the Auger processes under consideration are strongly influenced by shake-up transitions. Direct multiple Auger processes  \cite{Mueller2015a} resulting from higher-order many-electron interactions \cite{Zhou2016} must be expected to be much weaker than the shake-up of valence electrons. Here we neglect this non-sequential multiple autoionization and only include shake-up processes in our calculations. This assumption seems justified as the results obtained in our shake-up picture agree well with the experiment. Furthermore, we also neglect fluorescent losses. The rates for radiative transitions in nearly-neutral atoms (here $\sim6\times10^{11}$~s$^{-1}$) are typically orders of magnitude weaker than the rates of the strong Auger transitions ($\sim2.5\times10^{14}$~s$^{-1}$) that contribute to the cascade processes. If a significant amount of radiative losses was neglected, the calculations would overestimate the cross sections for the charged reaction products. Consequently, the population of the low charge states, here mostly neutral oxygen, would be underestimated.

We compute the wave functions by utilizing the MCDF method as implemented in the GRASP  \cite{Joensson2007} and RATIP codes \cite{Fritzsche2012a}. The latter is applied to compute all the radiative as well as Auger transition rates. To account for the relaxation of the electron density and the shake transitions of the valence electrons, the nonorthogonality of atomic orbitals is considered through the application of the bi-orthonormal transformation \cite{Olsen1995}. We also assume that the individual Auger processes in the cascades are independent from one another and can thus be combined in a subsequent step in order to compute the branching fractions and the decay paths that contribute to a given $m$-electron detachment process.

Ab-initio calculations have been carried out for the
\begin{eqnarray}
\label{eq:seq_exc}
\mathrm{O}^-(1s^2\,2s^2\,2p^5 \; {}^2P) +  h\nu  &\to&  \mathrm{O}(1s \,2s^2\,2p^6 \; {}^2S)\label{eq:res}
\end{eqnarray}
resonance to predict its width, the branching fraction to different final charge states as well as an estimate for the expected absolute cross sections. The resonance width is dominated by the multiple Auger emission
\begin{eqnarray}
\label{eq:seq_Auger}
\mathrm{O}^-(1s\,2s^2\,2p^6)   & \to&  \mathrm{O}^{(m-1)+}(1s^2\,2\ell^{7-m}) + m \, e^-
\end{eqnarray}
where $m = 1,2,3$. Clearly, detachment of three electrons ($m = 3$) is energetically forbidden when just normal Auger processes are taken into account since, then, only low-lying levels below the double-ionization threshold are populated in neutral oxygen after the first electron emission. To better understand the observed multiple electron detachment, we performed a series of calculations where progressively more shake-up transitions were added to the Auger cascade (Eq.~\ref{eq:seq_Auger}). The simplest calculation (model A) does not include any shake-up transitions and hence there is no triple detachment predicted. Model B extends this by adding single $2s\to 3s$ and $2p\to 3p$ shake-up transitions. In subsequent models, we also include double excitations $2p^2\to 3s^2$ and $2p^2\to 3p^2$ (model C), $2s^2\to 3s^2$ (model D),  $2s^2\to 3p^2$ (model E), and  $2p^2\to 3d^2$ (model F) with each model including all excitations of the preceding ones. Table~\ref{tab:shake} shows the calculated resonance widths and branching fractions of the lowest charge states for these different models and compares them with the experimental branching ratio for singly and doubly-ionized oxygen (last row). Here, the letter given in the first column is used to refer to the models introduced above.

\begin{table}[ttt]
\caption{\label{tab:shake}Convergence of the resonance width and of the branching fractions for the production of neutral oxygen atoms and O$^{+}$ and O$^{2+}$ ions upon photodetachment of O$^-$ via the $1s\,2s^2\,2p^6\;^2S_{1/2}$ resonance as progressively more shake-up transitions are included in the present calculations. The last column gives the ratio of the numbers from the two preceding columns. For the explanation of the models A--F see text.
}
\begin{ruledtabular}
\begin{tabular}{ccdddd}
&  Width  & \multicolumn{3}{c}{Branching Fraction} & \multicolumn{1}{c}{Ratio} \\ \cline{3-5}\rule[0mm]{0mm}{4mm}
  Model & (meV) & \multicolumn{1}{c}{O} & \multicolumn{1}{c}{~~~O$^+$} & \multicolumn{1}{c}{~~~~~~O$^{2+}$} & \multicolumn{1}{c}{O$^{+}$/O$^{2+}$}  \\
\hline\rule[0mm]{0mm}{4mm}
  \!\!A & 133 & 0.77  & 0.23 & 0.0      &    \multicolumn{1}{c}{---}   \\
  B     & 131 & 0.78  & 0.22 & \sim 0.0 		&    \multicolumn{1}{c}{---}   \\
  C     & 153 & 0.64  & 0.36 & 0.004    &   106   \\
  D     & 161 & 0.56  & 0.42 & 0.016    &    26   \\
  E     & 174 & 0.46  & 0.48 & 0.059    &     8.1 \\
  F     & 166 & 0.51  & 0.44 & 0.042    &    10.6 \\
  \hline\rule[0mm]{0mm}{4mm}
  exp   & $164 \pm 14$ & \multicolumn{1}{c}{---} & \multicolumn{1}{c}{---}  &  \multicolumn{1}{c}{---}   & 10.3 \pm 0.8
\end{tabular}
\end{ruledtabular}
\end{table}

When shake-up transitions of a single $2s$ or $2p$ electron to the $n = 3$ shell are considered, the formation of O$^{2+}$ becomes energetically allowed but still remains negligible.
Similarly, the decay width of the $1s-2p$ resonance is also much lower than observed, when no or only single excitations are considered. The inclusion of double excitations in our models leads to a significant population of doubly-ionized final states and also significantly increases the decay width. In particular, all models that include double excitations yield a decay width that is in agreement with the experimental result $164 \pm 14$~meV (Fig.~\ref{fig:resonance}). Extension of the models to also include double excitations from the $2s$ shell does not change the decay width.

In contrast, the formation of doubly-ionized oxygen depends much more critically on the chosen shake model. Here, double excitations from the $2s$ shell, introduced in model D, are necessary to obtain a significant population of O$^{2+}$, which is partly counter compensated by excitations to the $3d$ shell. With the last three models (D-F), we obtain a contribution of triple detachment that is reasonably compatible with the experimental finding of $10.3\pm0.8$.

The inclusion of double excitations into our models is computationally extremely expensive so that, consequently, we could perform only limited studies regarding the contribution of shells with $n>3$. Therefore, we restricted our theoretical model to study single and double excitations of the form $2s\to ns$, $2p\to np$, $2p^2\to ns^2$, $2p^2\to np^2$ (model C). Within this restricted configuration expansion, the inclusion of excitations to the $n=4$ shell did not have a significant influence on the computed resonance width but led to a change of the O$^{2+}$ branching fraction from 0.004 (Tab.~\ref{tab:shake}) to 0.007. This is a rather minor effect, that is not able to explain the large abundance of doubly-ionized decay products. Additional inclusion of excitations to the $n=5$ shell did not produce any further significant changes. This shows that the incorporation of additional $2\ell^2 \to 3\ell'^2$ transitions by the models D, E, and F is more important than the consideration of shake-up transitions to shells with $n>3$.

The energy and transition rate for the resonant excitation (Eq.~\ref{eq:seq_exc}) were computed by adding four correlation layers to the initial approximation. This yields a well converged excitation energy of 525.3~eV very close to the experimental value of $525.6\pm0.1$~eV (Fig.~\ref{fig:resonance}). For this resonance, the absorption strength  can readily be calculated from the radiative transition rate and converges to a numerical value of $\bar{\sigma}_\mathrm{abs} = 5.0$~Mb~eV.

The resonance strengths in the double and triple detachment channels are obtained by multiplying the computed absorption resonance strength with the branching fractions as obtained from our cascade model shown in Tab.~\ref{tab:shake}. Our best theoretical estimate (model F) yields 2.20~Mb~eV and 0.21~Mb~eV, respectively, that are a factor of 2.3 larger than the corresponding experimental results. Up to the present, the reason of this quite large discrepancy remains unclear. Generally, we expect the branching fraction of O$^{2+}$ to be underestimated since the contributions of shells with $n > 3$ and direct double Auger decays are neglected. However, the non-sequential emission of two or more electrons is expected to mostly contribute to the decay of negative oxygen, whose decay width is in very good agreement with experiment. We note that a similar, not yet understood discrepancy between theoretical and experimental resonance strengths still remains for the K-shell detachment of B$^-$ \cite{Berrah2007a} and the $K$-shell ionization of O$^+$ \cite{Bizau2015}. Closer inspection of the electron emission from the doubly excited intermediate states as recently performed for the photoionization of Xe$^{5+}$ ions \cite{Bizau2016} may in the future shed more light on this remaining discrepancy.

In summary, we have measured \emph{absolute} cross sections for the (multiple) detachment of negatively charged O$^-$ ions. Apart from the double photodetachment, that was considered for other light ions  in previous studies \cite{Kjeldsen2001a,Berrah2001,Berrah2007a,Gibson2003a,Walter2006a,Gorczyca2004}, we here also report the \emph{triple} photodetachment in the photon energy range of $K$-shell excitation and ionization. Two $K$-shell ionization thresholds were identified and the corresponding threshold energies were determined. At higher energies, the sum of the double and triple detachment cross sections agrees with the measured absorption cross sections for neutral oxygen \cite{McLaughlin2013} as well as for  O$^+$ and O$^{2+}$ \cite{Bizau2015}. This suggests that the production of neutral oxygen by nonresonant $K$-shell detachment of O$^-$ plays only a minor role. The natural line width of the prominent $1s\,2s^2\,2p^6\;^2S$ photo-detachment resonance which occurs below the $K$-shell ionization threshold was determined by a separate high-resolution measurement. The present large-scale ab-initio calculations of resonant photodetachment include a (very) large number of deexcitation pathways in order to properly incorporate the shake processes of the valence electrons as well as the rearrangement of the electron density in the course of the autoionization. It appears that autoionization accompanied by double shake-up has to be taken into account for reproducing the major part of the experimental findings. Similar four-electron Auger processes have been discussed recently also for photoionization \cite{Mueller2015a, Zhou2016} and photorecombination \cite{beilmann2011a} of positively charged atomic ions. $K$-shell photodetachment of negative ions continues to be a challenge for state-of-the-art atomic theory. The implementation of the photon-ion merged-beams method at the world's brightest 3rd-generation synchrotron light source has opened the door to further explorations of this fundamental atomic process with even heavier atomic species.

This research was carried out at the light source PETRA\,III at DESY, a member of the Helmholtz Association (HGF). We would like to thank G.~Hartmann, F.~Scholz, and J.~Seltmann for assistance in using beamline P04, S.~Klumpp for his continuous support of the PIPE setup as well as B.~Ebinger and A.~Perry-Sa{\ss}mannshausen for the off-site determination of the ion-detection efficiency.  This research has been funded in part by the German Federal Ministry for Education and Research (BMBF) under contracts 05K10RG1, 05K10GUB, 05K16RG1, 05K16GUC, and 05K16SJA.


%
\end{document}